# Multimodal machine learning for materials science: composition-structure bimodal learning for experimentally measured properties

Sheng Gong[1]*, Shuo Wang[1]*, Taishan Zhu[1], Yang Shao-Horn[1,2], and Jeffrey C. Grossman[1]

[1]Department of Materials Science and Engineering, Massachusetts Institute of Technology, Cambridge, MA 02139, USA

[2]Department of Mechanical Engineering, Massachusetts Institute of Technology, Cambridge, MA 02139, USA

* These authors contribute equally

# Abstract

The widespread application of multimodal machine learning models like GPT-4 has revolutionized various research fields including computer vision and natural language processing. However, its implementation in materials informatics remains underexplored, despite the presence of materials data across diverse modalities, such as composition and structure. The effectiveness of machine learning models trained on large calculated datasets depends on the accuracy of calculations, while experimental datasets often have limited data availability and incomplete information. This paper introduces a novel approach to multimodal machine learning in materials science via composition-structure bimodal learning. The proposed **CO**mposition-**S**tructure Bimodal **Net**work (COSNet) is designed to enhance learning and predictions of experimentally measured materials properties that have incomplete structure information. Bimodal learning significantly reduces prediction errors across distinct materials properties including Li conductivity in solid electrolyte, band gap, refractive index, dielectric constant, energy, and magnetic moment, surpassing composition-only learning methods. Furthermore, we identified that data augmentation based on modal availability plays a pivotal role in the success of bimodal learning.

# Introduction

Recently, multimodal machine learning models, such as GPT-4(*1*), have profoundly transformed the influence of artificial intelligence on society(*2*). By definition, multimodal machine learning obtains information from different modalities, such as text, image, and audio(*3*), and fuses all the information for downstream tasks. Multimodal machine learning has attracted increasing attention in many research communities such as computer vision and natural language processing, because in many cases multimodal machine learning outperforms the single modal learning(*4*). There also exist many modalities in materials science, such as composition, structure, spectrum, image, and text(*5*), which in principle can be simultaneously incorporated into multimodal machine learning models that potentially outperform machine learning models trained on single modality. Despite the potential, however, there is still a lack of widespread application of multimodal machine learning in the field of materials science. Therefore, it is important to demonstrate the effectiveness of multimodal machine learning in materials science.

One of the ultimate goals of materials informatics is to predict materials properties that are close to experimental measurements(*6*). Although most of current machine learning models applied to materials are trained on theoretical datasets(*7-16*) due to the high availability, the inherit gap between theoretical calculations and experiments indicates that the usefulness of such machine learning models largely depends on the accuracy of the theoretical calculations(*15, 17-19*). To bypass the dependency of accuracy of training data, recently experimentally measured properties, the "ground truth", have been used as the training set for machine learning models(*12, 20, 21*).

However, learning experimental data faces many challenges, such as limited number of data points, conflicted results in different literatures, and incomplete information recorded in literatures. While data cleaning has been used to assess the issue of conflicts(*22*), and machine learning techniques such as transfer learning(*12, 21, 23, 24*) and multi-fidelity learning(*20*) have been developed to alleviate the issue of limited number of data points, the challenge of incomplete information has not yet been addressed properly. This is because, single modal machine learning cannot encode all information. For example, as illustrated in Figure 1a, in most materials datasets the composition information of most materials is available(*25, 26*), but the data points with available structure information is a subset of data points with composition(*22, 25*). Therefore, although structure is more informative than composition(*27*), machine learning models trained on structures have to be trained on subsets. On the other hand, machine learning models trained on composition with more data cannot incorporate structure information into the model. As a result, available single modal machine learning models trained on either composition or structure are not ideal models that fully utilize all available information.

In this work, we propose that, multimodal machine learning, specifically composition-structure bimodal learning, can be used to improve the learning performance of experimentally measured materials properties. We build a new machine learning framework, COSNet, to realize a **CO**mposition-**S**tructure bimodal **Net**work. We show that composition-structure bimodal learning has better predictions for various materials properties than composition-only learning, including Li conductivity in solid electrolyte, band gap, refractive index, dielectric constant, energy, and magnetic moment. The improvement from addition of structure information exists in both data points with and without structures, showing the effect of representation alignment that the addition of the structure modality has globally informed the representation of the composition modality. We also find that the modal availability-based data-augmentation is critical to the effect of bimodal learning, as it ensures that the composition network is effectively trained by all available data.

The main contributions of the work are:

1) proposing the composition-structure bimodal learning framework to improve the learning performance to datasets with incomplete structure information.

2) demonstrating that data augmentation is critical to the improvement from bimodal learning.

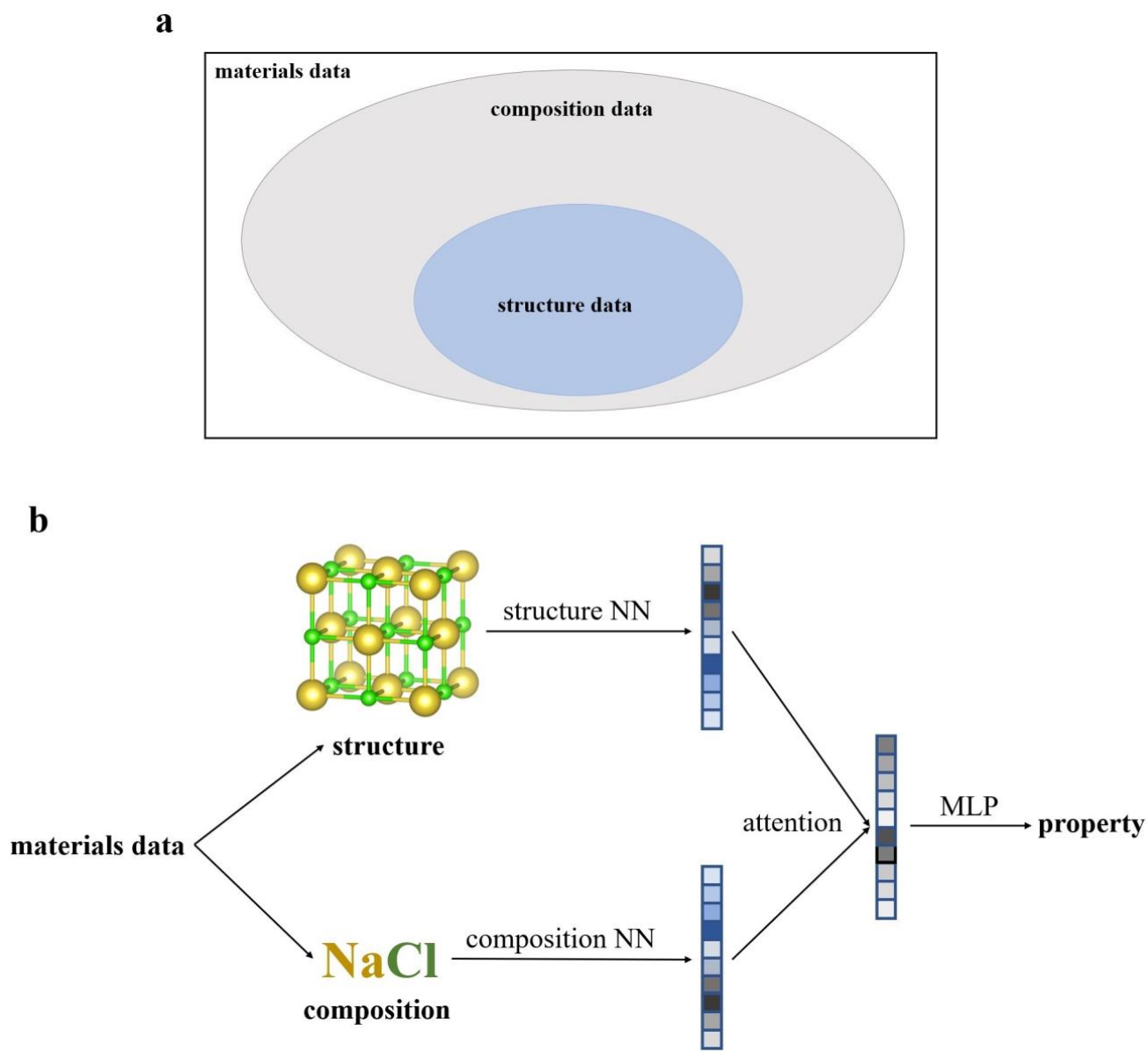

**Figure 1. Overview of composition-structure bimodal learning. a** Illustration of the range of materials data, composition data and structure data. Composition data is a subset of all materials data, and structure data is a subset of composition data. **b** Schematic of composition-structure bimodal learning and the COSNet.

## Results

**Architecture of COSNet.** In this work, we propose COSNet (**CO**mposition-**S**tructure Bimodal **Net**work) for composition-structure bimodal learning on materials dataset. As shown in Figure 1b, COSNet is mainly composed of four parts: a neural network to encode composition, a neural network to encode structure, an attention-based summation/concatenation to combine composition representation and structure representation, and a multi-layer perceptron (MLP) to predict property based on the combined representation.

The architecture of COSNet is summarized below. For a materials dataset $D = \{(c^i, s^i, t^i)_{i=1,...,N}\}$, where $c^i, s^i, t^i$ represent composition, structure, and target property of material $i$, respectively, composition representation $C^i$ and structure representation $S^i$ are first obtained by the composition network $g_C^r$ and the structure network $g_S^r$ as in equation (1) and (2), respectively:

$$C^i = g_C^r(c^i) \ \ \ ...... (1),$$

$$S^i = g_S^r(s^i) \ \ \ ...... (2),$$

In this work, both $g_C^r$ and $g_S^r$ are graph neural networks (ROOST(*10*) and de-CGCNN(*28, 29*), respectively). Note that, if material $i$ does not have recorded structure, we assign a null structure to it as a place holder: $s^i = s^{null}$. Then, to combine the two representations from the two modalities into an overall materials representation $M^i$, we design an attention-based summation/concatenation as below:

$$M^i = \begin{cases} C^i * w_c^i + S^i * w_s^i \\ \quad \quad \text{or} \\ C^i * w_c^i \oplus S^i * w_s^i \end{cases} ...... (3),$$

where $+$ is element-wise summation and $\oplus$ is vector concatenation, and $w_c^i$ and $w_s^i$ are the attention weights for $C^i$ and $S^i$, respectively. Note $w_c^i$ ($w_s^i$) can be either scalar or vector with the same dimension of $C^i$ ($S^i$), and this choice of dimension of attention weight and the choice between $+$ and

$\oplus$ are defined as hyper-parameters in COSNet. $w_c^i$ and $w_s^i$ are determined as below:

$$w_c'^i = \text{MLP}_C^w(C^i) \quad \ldots\ldots (4),$$

$$w_s'^i = \text{MLP}_S^w(S^i) * p_S^i \quad \ldots\ldots (5),$$

$$w_c^i = \frac{w_c'^i}{w_c'^i + w_s'^i} \quad \ldots\ldots (6),$$

$$w_s^i = \frac{w_s'^i}{w_c'^i + w_s'^i} \quad \ldots\ldots (7),$$

$$p_S^i = \begin{cases} 0, & \text{if } s^i = s^{null} \\ 1, & \text{if } s^i \neq s^{null} \end{cases} \quad \ldots\ldots (8),$$

where $\text{MLP}_C^w$ and $\text{MLP}_S^w$ denote the multi-layer perceptron (MLP) for determining composition and structure weights, respectively. Note that a softplus activation function is used at the end of $\text{MLP}_C^w$ and $\text{MLP}_S^w$ to ensure that $w_c'^i$ and $w_s'^i$ are positive. Combining equations (3) to (8), the materials representation $M^i$ can be written as:

$$M^i = \begin{cases} C^i \text{ or } C^i \oplus \mathbf{0}, & \text{if } s^i = s^{null} \\ C^i * w_c^i + (\text{or } \oplus) S^i * w_s^i, & \text{if } s^i \neq s^{null} \end{cases} \quad \ldots\ldots (9),$$

where $\mathbf{0}$ is a zero vector with the same dimension of $S^i$. Finally, a MLP is used to predict target property from $M^i$:

$$t^i = \text{MLP}^t(M^i) \quad \ldots\ldots (10).$$

Besides the neural network architecture of COSNet, the other critical aspect of composition-structure bimodal learning in this work is the data augmentation based on the relation between composition and structure. For a composition $c$, in principle there are infinite number of structures $s$ that have the same composition $c$. If we impose a restriction that $s$ should only include stable or meta-stable structures at room temperature and pressure, then there might be a single or multiple $s$ corresponding to a given $c$, depending on the nature of the composition $c$. In a materials dataset $D$, for a given $c$, there might be no or a single or multiple $s$ that have composition $c$. If only one specific $s^i$ in the dataset has the composition $c^i$, then ideally the prediction from composition-structure bimodal learning should be equal to that from composition-only learning:

$$\forall c^i \in D \text{ s.t. } \exists! \; s^i \in D, f_C(s^i) = c^i,$$

$$\text{MLP}^t(C^i \text{ or } C^i \oplus \mathbf{0}) = \text{MLP}^t(C^i * w_c^i + (\text{or } \oplus) \; S^i * w_s^i) \; \ldots \ldots (12),$$

where $f_C$ denotes the mapping function of a structure $s$ to its composition $c$. However, different from the invariance imposed by physics such as translation and rotation invariance of structure-property learning(*28, 30*), the equality in equation (12) is dataset-dependent and might break when the dataset is changed, such as when a new structure $s^j$ corresponding to an existing $c^i$ is added into the dataset. Therefore, it is not reasonable to impose equation (12) in the neural networks of COSNet. Instead, we implement a data augmentation in the data preparation stage of COSNet for equation (12):

$$D_{\text{augmented}} = \{(c^i, s^i, t^i)_{i=1,\ldots,N}\} + \{(c^i, s^{null}, t^i)_{\forall (c^i, t^i) \in B}\},$$

$$B = \{(c^i, t^i), \exists! \; s^i \in D \text{ s.t. } f_C(s^i) = c^i\} \; \ldots \ldots (13).$$

In other words, if a composition only corresponds to one structure in the training set, then this composition is added into the training set without structure, which is to push COSNet to have similar predictions based on only composition and the pair of composition-structure. Note that the data augmentation is only implemented on training sets, not on validation sets and test sets.

**Predictions from COSNet.** In Table 1, we compare the predictions from composition-structure bimodal learning (COSNet) and composition-only learning (COSNet with $\forall i \in D, p_s^i = 0$) for four experimentally measured materials properties: Li conductivity in solid electrolyte, band gap, refractive index, and dielectric constant. We can see that, for the four properties, bimodal learning has better predictions for all the complete datasets than composition-only learning with around 7% to 10% lower mean absolute errors (MAEs), and the improvements are statistically significant as the difference of errors are larger than the sum of the standard deviations of the two MAEs. We can also see that, the lower errors of bimodal learning are observed for both data points with structures and without structures, which shows that structure information of a portion of data points can improve prediction of other data points without structure.

Table 1. Comparison of test set mean absolute errors (MAEs) of composition-only learning, bimodal learning with data augmentation and bimodal learning without data augmentation, respectively. For each property, the top, middle, and bottom row represent MAE of all data in the test set, data with only composition, and data with both composition and structure, respectively. The unit of each property is shown under the name of the property.

| Model | dataset type | number of data points | composition only | bimodal with augmentation | bimodal without augmentation |
|---|---|---|---|---|---|
| Li Conductivity ($\log_{10}$ (mS/cm)) | complete | 1499 | 1.022 ± 0.047 | **0.924 ± 0.012** | 1.245 ± 0.078 |
| | w/o structure | 1077 | 0.906 ± 0.048 | **0.832 ± 0.033** | 1.122 ± 0.099 |
| | w/ structure | 422 | 1.425 ± 0.091 | **1.243 ± 0.116** | 1.674 ± 0.145 |
| Band Gap (eV) | complete | 5901 | 0.502 ± 0.016 | **0.460 ± 0.004** | 0.499 ± 0.009 |
| | w/o structure | 2741 | 0.435 ± 0.031 | 0.415 ± 0.009 | **0.404 ± 0.008** |
| | w/ structure | 3160 | 0.560 ± 0.009 | **0.495 ± 0.010** | 0.577 ± 0.009 |
| Refractive Index (1) | complete | 1736 | 0.473 ± 0.008 | **0.430 ± 0.010** | 0.482 ± 0.006 |
| | w/o structure | 995 | 0.492 ± 0.007 | **0.421 ± 0.023** | 0.441 ± 0.009 |
| | w/ structure | 741 | **0.436 ± 0.015** | 0.439 ± 0.014 | 0.514 ± 0.009 |
| Dielectric Constant (1) | complete | 1768 | 235 ± 6 | **219 ± 7** | 225 ± 4 |
| | w/o structure | 1140 | 295 ± 12 | **291 ± 10** | 300 ± 6 |
| | w/ structure | 628 | 118 ± 16 | 72 ± 0 | **71 ± 1** |

Here we propose that the improvement of data points without structure is from representation alignment(*3*), which is a phenomenon that representations from different modalities ($C^i$ and $S^i$ in this work) align with each other during multi-modal training. In other words, $C^i$ and $S^i$ are aligned during the training, and since $S^i$ contains more information than $C^i$, $C^i$ becomes more informative in bimodal learning than composition-only learning, which leads to better predictions of data points without structure. To demonstrate the hypothesis, here we use a structural descriptor, volume per atom ($V_a$, a descriptor strongly correlated with ion conductivity(*31*)), to represent structure information, and use $C^i$ to predict $V_a$ by linear regression and take the $R^2$ scores of the linear regressions as a metric to evaluate correlation between $C^i$ and structure. In Table 2, we show that $C^i$ from bimodal learning has stronger correlation with $V_a$ than that from composition-only learning, which verifies the effect of representation alignment between $C^i$ and $S^i$ and explains why addition of structure information helps learning of data points without structure.

Table 2. $R^2$ scores of the linear regressions between $C^i$ and $V_a$ of composition-only learning, bimodal learning with data augmentation and bimodal learning without data augmentation, respectively. The dataset for the linear regressions is composed of data points in the test set of Li conductivity that have structure.

| Model | $R^2$ scores |
| --- | --- |
| composition-only | $0.373 \pm 0.034$ |
| bimodal with data augmentation | $0.435 \pm 0.048$ |
| bimodal without data augmentation | $0.371 \pm 0.019$ |

In addition to the four experimental datasets in Table 1, in Figure 2 we plot the MAE versus percentage of data points with structures for magnetic moment per formula and energy per atom. Although these two datasets are purely theoretical from the Materials Project database(*32*), we include them here to show the impact of percentage of data points with structures as well as how bimodal learning performs on larger datasets than typical experimental datasets as in Table 1. We can see that, for the larger theoretical datasets, bimodal learning also has better predictions than composition-only learning (percentage = 0), and the more structures, the lower errors, which demonstrates the positive effect of structure on predicting materials properties. The trend of lower errors with respect to more structures is inspiring, as it indicates that, in the future with more structures measured in the experimental datasets, more significant improvement can be achieved by bimodal learning compared with that in this work.

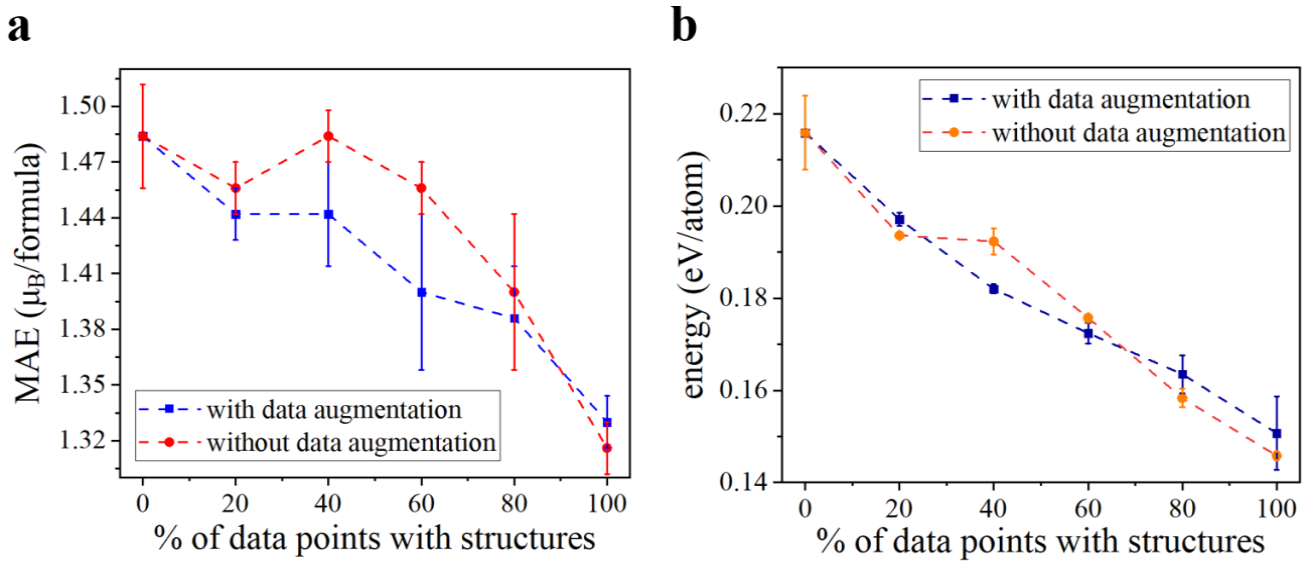

**Figure 2.** Test set MAE of COSNet versus percentage of data points with structures for total magnetic moment per formula and energy per atom. The two datasets are composed of 25,000 materials randomly sampled from the Materials Project database(*32*).

**Effect of data augmentation.** In addition to the comparison between bimodal learning and composition-only learning, we also compare bimodal learning with data augmentation in equation (13) and bimodal learning without data augmentation in Table 1 and Figure 2. We can see that, for all the complete datasets in Table 1, bimodal learning without data augmentation has larger errors than that with data augmentation, and for Li conductivity and refractive index, bimodal learning without data augmentation even has larger errors than composition-only learning. As in Figure 2, for moderate percentages, bimodal learning without data augmentation has larger errors than that with data augmentation. From Table 2, we can see that the effect of representation alignment is more significant on bimodal learning with data augmentation than that without augmentation, which is straightforward as data augmentation in equation (13) directly pushes $C^i$ and $C^i * w_c^i + (\text{or } \oplus) S^i * w_s^i$ to align with each other and result in similar predictions. The stronger representation alignment might explain why data augmentation leads to better predictions.

A possible reason for the poorer predictions and weaker representation alignment from bimodal learning without data augmentation is that, without data augmentation the structure network and the composition network might be trained separately. This is because, structure is generally more informative than composition for predicting properties(27). Therefore, for data points with structures, bimodal learning might underplay the role of the composition network. To support the argument, we plot the norm of $w_c^i$ for predictions of test set of Li conductivity. For data points without structure, both bimodal learning with and without augmentation solely use the composition network to predict the property with $w_c^i = 1$. However, for data points with structures, we can see that bimodal learning without augmentation has much smaller $w_c^i$ than bimodal learning with augmentation, and for more than a half of data points $w_c^i < 0.5$. Although structure has more information than composition, the number of data points with structures is usually very limited as in Table 1. Consequently, bimodal learning without structure somewhat splits the learning task into two subtasks, one for training the structure network and the other training the composition network, leading to the result that the composition network is not adequately trained. On the contrary, with data augmentation, the composition network is trained on the complete composition dataset, and the structure channel is also trained on all available structure data, which leads to better training and prediction.

The phenomenon in Figure 2 that, the differences between MAEs of bimodal learning with and without data augmentation are most significant at 40% and 60%, also supports the analysis above. At the moderate percentages, for bimodal learning without data augmentation, the composition network is not well trained, which leads to the largest difference compared with that with data augmentation where both networks are trained on all available data. However, at smaller percentages the composition network dominates and get trained by most data points without structure, and at higher percentages the structure network dominates and get trained by most data points with structures. Therefore, the difference between bimodal learning with and without data augmentation becomes less significant at small and high percentages of data points with structures.

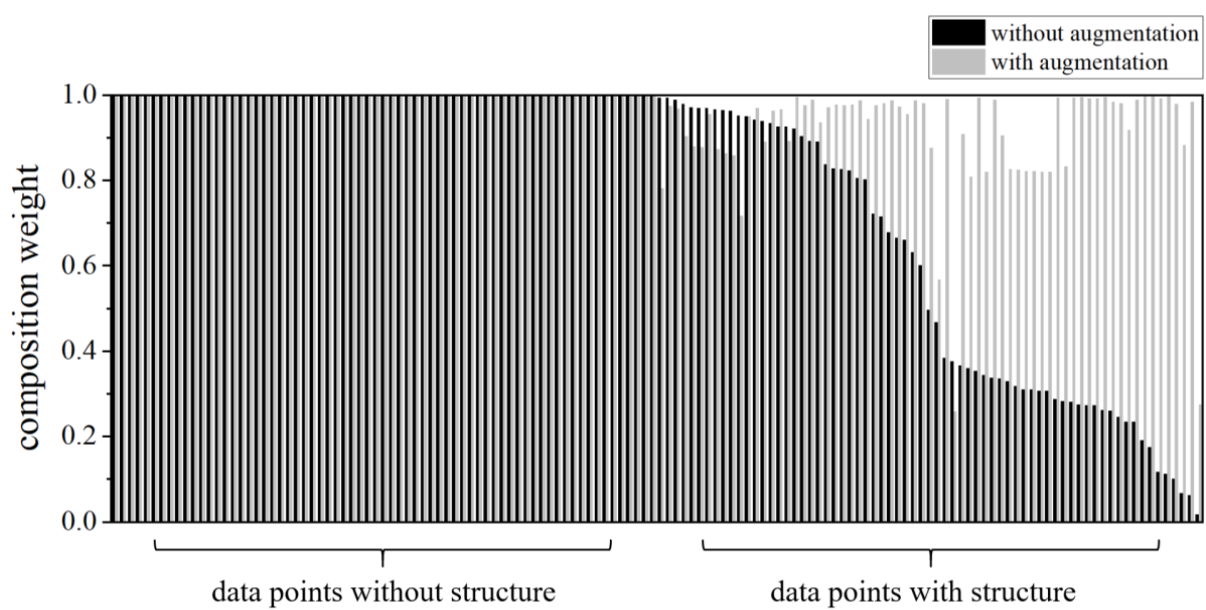

**Figure 3.** Visualization of composition weights for predicting Li conductivity. Note that the length of the two regions in the figure is not strictly scaled to the percentage of number of data points with and without structure.

## Discussions

In this work, we confront the challenge that, many materials datasets, especially experimental datasets, usually have complete composition information but incomplete structure information, which limits the learning performance of single modal machine learning models as a model trained on either composition or structure cannot incorporate all available information. We propose that both modalities of composition and structure can be used simultaneously to predict materials properties, and we design a machine learning framework, COSNet, to realize the composition-structure bimodal learning. Compared with composition-only learning, we find that bimodal learning can achieve lower errors on four experimental datasets (Li conductivity, band gap, refractive index, and dielectric constant) and two theoretical datasets (energy and magnetization). Moreover, these improvements exist in both data points with and without structures, which shows the effect of representation alignment between composition and structure. We also find that, data augmentation is critical to the improvement from bimodal learning, as data augmentation strengthens representation alignment and ensures that the composition network can be adequately trained by all available composition data.

Note that a large portion of structures used in this work are from the Materials Project database(*32*), which might not correspond to the true structures measured in the experiments. Therefore, with more reliable experimentally measured structures in the future, we expect that improvement from the bimodal learning for experimental properties can be more significant than this work. As above, this work establishes a new avenue for learning and prediction of experimentally measured materials properties, a key task in the field of materials informatics.

As a proof of concept of multimodal machine learning for materials science, we hope this study can inspire future studies further improving the machine learning architectures for multimodal learning, incorporating more modalities, and applying multimodal learning on more scenarios in materials science. With new machine learning models being developed for each single modality, such as new composition networks and new structure networks, we hope that the improvement from composition-structure bimodal learning can be more significant than this work. Beyond composition and structure, there are many other modalities in materials science, such as spectrums, images, and texts. We hope that future studies can simultaneously incorporate more modalities into machine learning models. With future development of multimodal machine learning for materials science, we hope that one day people can build up "materials general intelligence" that obtains information from all modalities of materials and solves tasks for different purposes of materials development, like what GPT-4 is doing today(*1*).

## Methods

**Datasets.** In this work, we use four experimental datasets and two theoretical datasets to test the idea of bimodal learning and COSNet. The dataset of Li conductivity in solid electrolyte is from Ref.(*33*) and Ref.(*34*), and a data cleaning is conducted that the value from the most recent report is used for conflicted entries. Structures in the dataset of Li conductivity are from the Materials Project database(*32*) with a manual match. The dataset of band gap is from Ref.(*35*) and the Springer Materials database(*36*). In the band gap dataset, the structures are selected from the Materials Project database(*32*) with available space group information, and for conflicted entries of band gap we use the mean value as the value in our dataset. The datasets of refractive index and dielectric constant are from Ref.(*26*), and a data cleaning is conducted that removes compositions of organic materials and replaces the conflicted values by the mean value. Structures in the two databases are manually selected from the Materials Project database(*32*) and the ICSD database(*37*) according to the available reference information in Ref.(*26*). The datasets of total magnetization per formula and energy per atom are randomly sampled from the Materials Project database(*32*). All the datasets are randomly split (60%:20%:20%) into training, validation and test sets for training and evaluation of the machine learning models.

**Models.** In this work, we choose ROOST(*10*) and de-CGCNN(*28, 29*) as the composition and structure network in COSNet, respectively. This choice is based on the fact that, as in Table 1, experimental datasets typically have limited number of data points, and both ROOST(*10*) and de-CGCNN(*29*) have been shown to have strong ability for small datasets. All the models are trained by 200 epochs, and the epoch that has the lowest validation MAE is selected as the final checkpoint for each training. Hyper-parameters search is conducted according to Table 3 to find the hyper-parameters with the lowest validation error, and the selected model is used to predict the test set as the final evaluation. The training-validation-test procedure is repeated five times with different random seeds to report the error bars in Table 1, Table 2 and Figure 2.

Table 3. Hyper-parameters search space for COSNet in this work. Parameters not mentioned here are set to the default value as in the open source codes.

| Name | Space |
| --- | --- |
| weight decay | 1e-1, 1e-2, 5e-3, 1e-3, 5e-4, 1e-4 |
| learning rate | 1e-2, 5e-3, 1e-3, 5e-4, 1e-4 |
| learning rate decay | 0.99, 0.98, 0.97, 0.96, 0.95, 0.94 |
| concatenation | false, true |
| scalar weight | false, true |

**Funding:** This work was supported by Toyota Research Institute.

**Author Contributions:**

Conceptualization: SG, SW, YSH, JCG

Methodology: SG, SW

Investigation: SG, SW

Supervision: YSH, JCG

Writing—original draft: SG, SW, YSH, JCG

Writing—review & editing: SG, SW, YSH, JCG

**Competing Interests**: The authors declare no competing interest.

**Acknowledgement:** We thank Dr. Kiarash Gordiz and Daniele Vivona for helping collect the Li conductivity dataset.

**Data and Materials Availability:**

All datasets and codes in this work are provided at:

https://github.com/shenggong1996/COSNet/tree/master

Crystal structures and materials properties from the Materials Project database (V2021.03.22) are downloaded at https://materialsproject.org/.